# A Bandwidth Characterization Tool For MPEG-2 File

Sandeep. Kugali, A. V. Sutagundar
Basaveshwar Engineering College Bagalkot
Karnataka, India
intel_san@yahoo.co.in, ashok_ec@yahoo.com

S. S. Manvi
Wireless and Information Research Laboratory
Reva Institute of Technology and Management
Bangalore, India
agentsun2002@yahoo.com

*Abstract*— **This paper proposes the design and development of Moving Picture Experts Group-2 (MPEG-2) Video Decoder to offer flexible and effective utilization of bandwidth services. The decoder is capable of decoding the MPEG-2 bit stream on a single host machine. The present decoder is designed to be simple, but yet effectively reconstruct the video from MPEG-2 bit stream. The designed MPEG-2 video decoder will display the name of each frame and corresponding frame bandwidth and also displays the buffer fullness for each picture, bit rate, time interval, maximum quantization error and decoded size of the pervious picture etc,. In order to characterize the bandwidth requirement, the proposed multimedia tool is tested on various MPEG-2 bit streams for the operation of effectiveness in terms of various performance parameters such as: cumulative bandwidth, peak bandwidth, average bandwidth and minimum bandwidth.**

## I. INTRODUCTION

Multimedia is often used in reference with creativity using the PC. In simple words Multimedia [1] means multi (many) and media (communication/transfer medium). Hence it means many mediums working together or independently. It can be considered to be a combination of several mediums: **Text, Graphics, Animation, Video and Sound.** They all are used to present information to the user via the computer. The main feature of any multimedia application is its human interactivity. Each frame of a video is represented by a fluctuating voltage and is stored to a media commonly known as VHS cassette tape. All of the video components (color and the luminance signal) are combined into one signal [2].

Digital transmission generally represents the signals as voltages that alternate between two possible states (yes/no, on/off or 1/0). This is a base band signal [3]. It is difficult to determine the required bandwidth [4] just from the user's preference on perceived video quality. A typical multimedia application needs 1-10 Megabits per second bandwidth to enable communication in a continuous and interactive fashion.

Compression is a reversible conversion (encoding) of data that contains fewer bits. These compression algorithms often use statistical information to reduce redundancies. Huffman-Coding [5] and Run Length Encoding [6] are two popular examples allowing high compression ratios depending on the data.

In this Paper, the system model MPEG-2 Video Decoder reconstructs the video from MPEG-2 bit streams by taking the main parameters of Bandwidth into an account. The model mainly consists of Variable Length Decoder (VLD), Inverse Scan, Inverse Quantization, Inverse Discrete Cosine Transform (IDCT), frame memory and Motion Compensation (MC) based prediction modules.

Some of the related works are as fallows: The work given in [7], proposes an implementation of the MPEG-2 video decoder on the Digital Signal Processor (DSP), which implements all the MPEG-2 main-profile-at-main-level functionality, and conforms to the expressDSP™ Algorithm Standard (xDAIS) to enhance reusability. In [8], A Single-Chip MPEG-2 Video Decoder Large Integrated Circuit (LSI) is designed. Some other related work details regarding MPEG-2 are elaborated in [9] [10] [11] [12].

The rest of the paper is organized as follows. Section II presents a brief background of MPEG-2 theory. Section III presents the system model. Section IV presents the results and discussion.

## II. MPEG-2 THEORY

MPEG-2 [10], [11] is an extension of the MPEG-1 video-coding standard. MPEG-1 was designed to code progressively scanned video at a bitrate of 1.5 Mbps. On the other hand MPEG-2 is directed at broadcast formats at higher data rates of 4 Mbps Digital Versatile Disc (DVD) and 19 Mbps High Definition Television (HDTV).

### A. MPEG-2 Video Bit stream

The MPEG-2 bit stream is divided into several layers in a hierarchy which is as shown in figure 1.

**Sequence:** Contains a sequence header containing information such as horizontal and vertical size of the pictures, bitrate and chroma format.

**Group of pictures (GOP):** Used for skipping the sequence or random access of the sequence, when the instance of jumping to a certain time code. The header contains time information.

**Picture:** Contains information about the picture structure, scan type and other flags needed in the decoding process.

**Slice:** A series of consecutive macroblocks. A slice must start and end in the same horizontal row. Used for error concealment, if a bit stream error occurs the decoder should look for the next slice. Header only contains a quantization scale code.

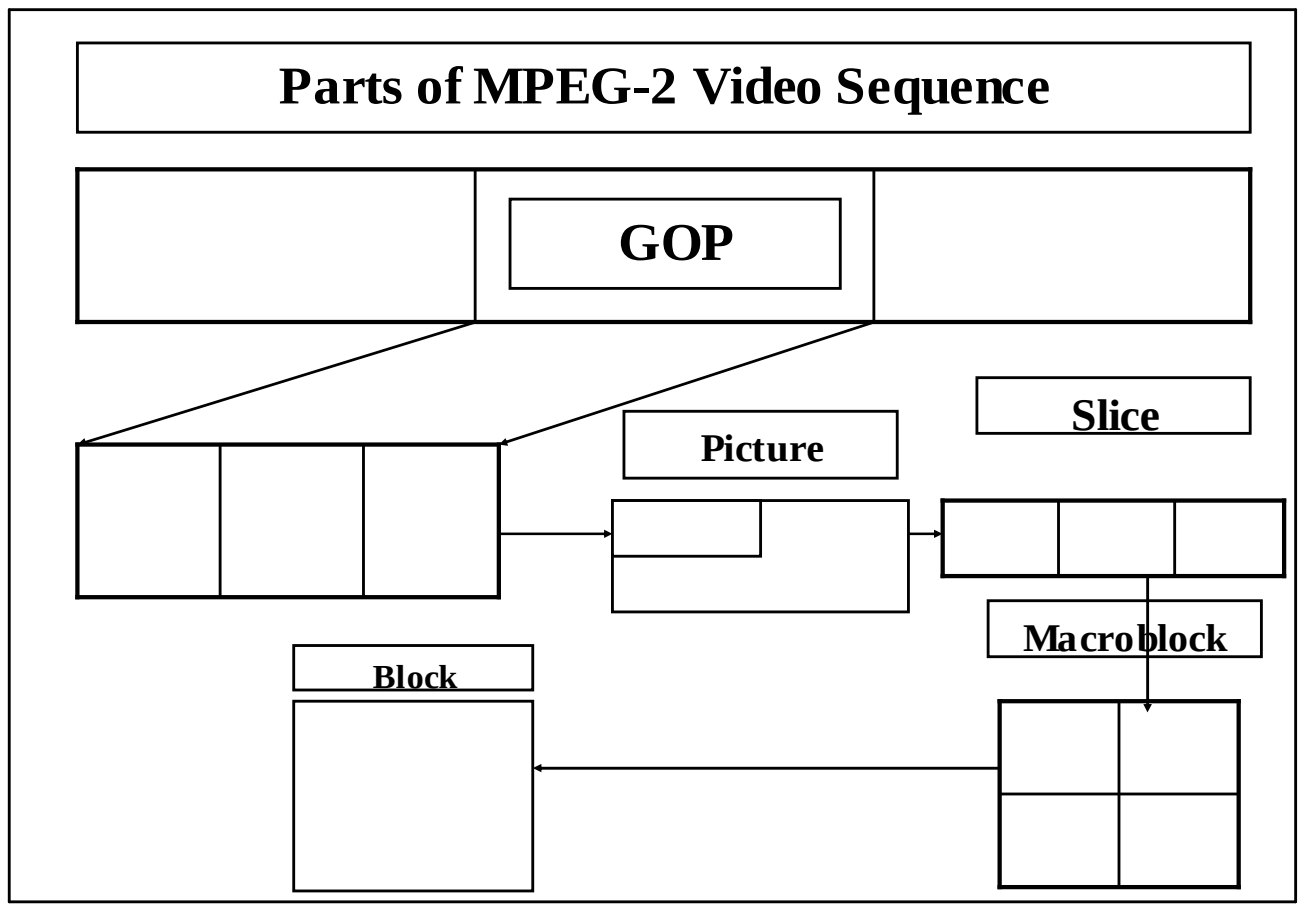

Figure 1: Overview MPEG-2 bit stream structure

**Macroblock:** In the case of 4:2:0 chroma subsampling format, a macroblock contains a header and up to 4 blocks of Luminance (Y) components and 2 blocks of Chrominance blue difference (Cb) and Chrominance red difference (Cr) components. The data in a macroblock is for the most part Huffman coded as explained in the next section. The macroblock header contains an address increment, motion vectors and a pattern code. The address increment is used for macroblock skipping. The pattern code is a 6 bit flag that indicates which of the blocks are available in the stream.

**Block:** Is an 8x8 Huffman and Discrete Cosine Transform (DCT) coded pixel block which is used in the most part of intra-pictures.

### B. *MPEG-2 Video Encoder*

In the MPEG-2 bit stream, start codes are used to signal the start of the different parts of the bit stream [13]. The basic block diagram of an MPEG-2 video encoder is as shown in the figure 2.

**DCT:** The MPEG-2 encoder uses 8x8, 2-D DCT [14].

**Quantizer:** The DCT coefficients obtained above are then quantized by using a default or modified matrix. User defined matrices may be downloaded and can occur in the sequence header or in the quantization matrix extension header. The quantizer step sizes for DC coefficients of the luminance and chrominance components are 8, 4, 2 and 1 according to the intra DC precision of 8, 9, 10 and 11 bits respectively. For the AC coefficients the corresponding header information to the quantizer scale factors are the quantizer scale type in the picture coding extension header and the quantizer scale code in the slice and Macro Block (MB) headers.

**Motion estimation and compensation:** In the motion estimation process, motion vectors for predicted and interpolated pictures are coded differentially between macroblocks.

**Scanning and VLC:** The quantized transform coefficients are scanned and converted to a one dimensional array. Two scanning methods are available: zigzag scan which is typical for progressive (non-interlaced) mode processing, and alternate scan which is more efficient for interlaced format video. The list of values produced by scanning is then entropy coded using a variable length code (VLC). Huffman coding is an entropy coding scheme for data compression.

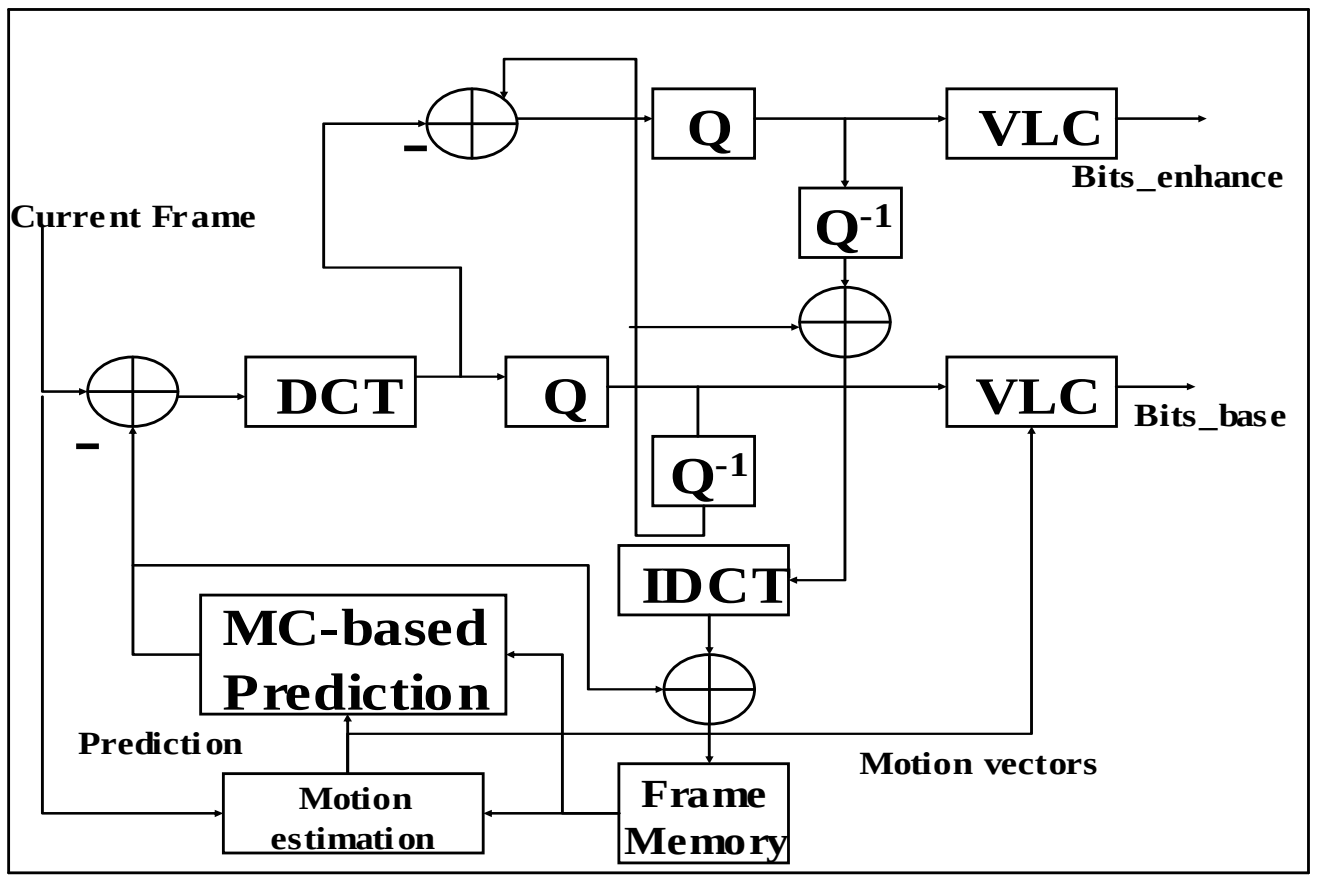

Figure 2: MPEG-2 Video Encoder

## III. System Model

In this section, the system model (MPEG-2 Video Decoder) which is as shown in figure 3 and its algorithms are discussed.

### A. *MPEG-2 Video Decoder*

**Decode Headers:** This part extracts header data and decodes it. Each of the block checks the current bit stream for a certain start code. The start code is a 32 bit sequence where the first 23 bits are zero followed by a one and a specific start code. For instance the start code for the sequence header is “00 00 01 B3” in hexadecimal.

**Decode Macro block Mode:** This part extracts the macro block header and data, which are ordered in a structure with both Huffman coded data and non-coded data. The reason why the macro block header decoder does not belong to the header extractor is that it has no start code. The macroblock decoding will continue until a new header start code is found in the bit stream. Since the start code begins with 23 zeros it will naturally stop the macroblock header decoding, hence there are no Huffman codes with that many zeros in row. Once a slice header is found in the header extractor, a new reset will start the Macroblock (MB) header decoder again.

When the entire header is decoded the reset signal reaches the MB block data decoder and block decoding begins. The block pattern from the macroblock header indicates the number of blocks in the current macroblock. The MB block data decoder finds, either a DC value coded with a differential or a run length value used to build the pixel block.

**Calculate and add Motion Vectors:** The Buffer block is used for quick to macroblock skipping. When the address increment increases with more than one a macro-block is inserted in between the current macroblock and the previous. This is done by using default motion vector values and making predictions for the skipped blocks. Normally the implementation would require the decoding to be halted while the skipped blocks are predicted. The buffer stores the latest macroblock (header and data) while sending default skipped predictions to the motion vector prediction block. The output from the buffer is chrominance and luminance components. The macroblock header data is fed to the motion vector prediction block.

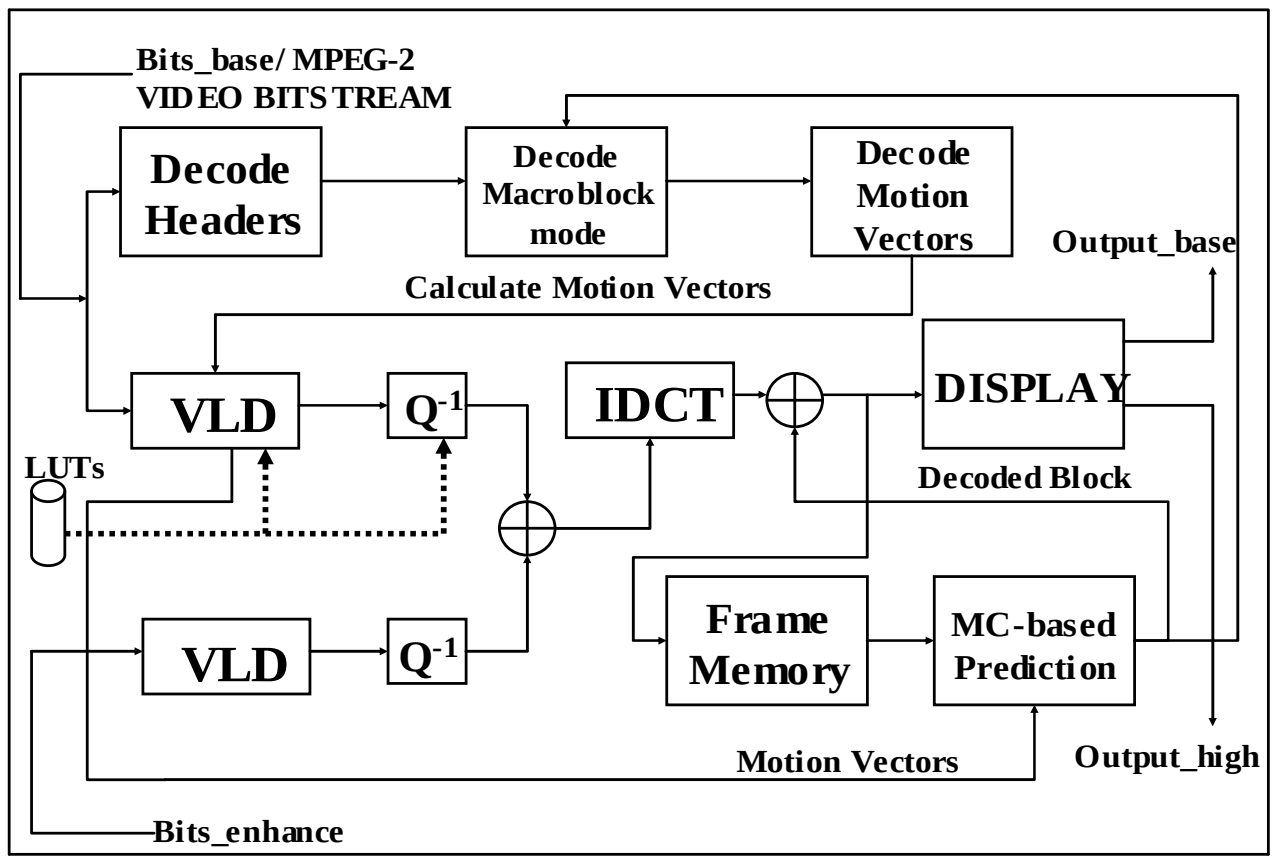

Figure 3: MPEG-2 decoder block diagram

**Variable length decoding:** It is a process that involves the use of table defined for decoding intra DC coefficients and also maintains three tables, one each for non intra DC coefficients, intra AC coefficients and non intra AC coefficients for decoding purpose. The output from the Huffman decoding is a 64-value vector. The first value in the vector is the most important one because it is the DC value of the DCT coded data. If the block contains the coefficients with zero value, then DC coefficient generates a pixel block where all the pixels have the same value. The DC value is kept as a predictor value, which will be reset to a default value when a new slice is found in the bitstream. Each intra-coded block will then add a differential to the predictor to get the current DC and produce a new predictor.

**Inverse scan:** The output of the variable length decoding stage is one-dimensional and of length 64. Inverse scan process converts one-dimensional data into a two dimensional array of coefficients according to a predefined scan matrix. The next step of the decoding process is to rearrange the 64-value vector to an 8x8 matrix. The alternate scan is meant to be used for interlaced blocks, where an odd and even line belongs to different parts of the picture. The scanning order is used to concentrate the lower index values from the 64-value vector to the top left corner of the matrix.

**Inverse quantization:** At this stage the two-dimensional DCT coefficients are inverse quantized to produce the reconstructed DCT coefficients. This process involves the rescaling of the coefficients by essentially multiplying them by the quantizer step size. The quantizer step size can be modified by using either a weighing matrix or a scale factor. After performing inversion quantization, the saturation and mismatch control operations are performed. Saturation is employed to keep the DCT coefficients in the range of [- 2048:+2047]. After the mismatch, control operation is performed to avoid large and visible mismatches between the state of the encoder and decoder. The mismatch control is done to correct an error that occurs due to the difference in the DCT at the encoder and the IDCT. The quantization is a more or less a low pass filtering process where the coefficients are rounded off to zero. If the block contains more zeros, Huffman coded bit string will be shorter.

**Inverse DCT:** After reconstructed DCT coefficients are obtained, a 2D 8x8 inverse DCT is applied to obtain the inverse transformed values. These values are then saturated to keep them in the range of [-256:+255]. In a not skipped macroblock, if the pattern code is one for a particular block in the macroblock then the corresponding coefficients for that block is included in the bitstream, else the pattern code is zero or the macroblock is skipped then that block contains no coefficient data. The 2D-DCT is used to transform the 8x8 pixel block from the spatial domain to the frequency domain. Since the energy in the frequency domain is mainly concentrated to the lower frequencies (at the top left corner of a matrix) and the higher frequencies have smaller components which can be rounded to zero by the quantization process without degrading the picture quality. The inverse process restores the pixels from the frequency domain to the spatial domain. However the 2D-IDCT can be done by doing two one dimensional IDCTs on the matrices, one on rows and one on columns.

**Motion Compensation:** In this stage, predictions from previously decoded pictures are combined with the inverse DCT transformed coefficient data to get the final decoded output. The prediction modes are the different ways a block of pixels is moved from the reference frame to the current. The prediction mode may be switch on a macroblock-to-macroblock basis. In field pictures, the actual picture is either a top or a bottom field picture and in frame pictures the picture has both fields as odd and even rows in the same picture. Therefore it is twice the size of a field picture.

**Frame Prediction, Frame Picture:** The simplest prediction mode is the frame prediction in frames of video, and it works in the same way regardless of the format of the image (interlaced or progressive). A single motion vector is used for the prediction of each forward or backward picture. The prediction is performed on 16x16 pixel matrix area and is transferred from the reference to the prediction. For an interpolated B-picture two motion vectors are used.

**Field Prediction, Frame Picture:** This mode is used for interlaced frame pictures. Since pictures have two fields from different points in time, it might in some cases be more efficient to use individual predictions for each field. Instead of predicting an entire 16x16 area, odd and even lines have separate predictions of 16x8 (eight rows). Each motion vector also has a flag which indicates from which field the prediction should be formed; both motion vectors may use the same field as reference. A B-picture with interpolated prediction will then use four motion vectors and four flags to form a prediction.

**Field Prediction, Field Picture:** The field prediction is the normal prediction mode for field pictures. A single motion vector used to do a 16x16 pixel area prediction. In field pictures the motion vector has a corresponding flag which corresponds top or a bottom field. If the prediction is done on the opposite field from the decoding and it will be in the same reference frame.

**Frame Store:** When a decoding of a picture is completed, depending on, whether it is an I- or P- picture, can be used as reference frame of the picture sampled, and is held as either a forward or backward reference frame. The frame store contains a block which points to the pictures in display order.

### *B. Algorithms of MPEG-2 Decoder*

MPEG-2 video decoding algorithm is mainly divided into following sub modules, which are discussed one by one. The sub-modules and corresponding titles are as shown in Table 1.

TABLE I. LIST OF SUB-MODULES WITH CORRESPONDING TITLES OF MPEG-2 VIDEO DECODER

| Sub-Module No. | Title |
|---|---|
| 1 | Bit-Level Routine |
| 2 | Header Decoding |
| 3 | Macroblock and Motion Vector Decoding |
| 4 | Variable Length Decoding |
| 5 | Inverse Quantization |
| 6 | Inverse Discrete Cosine Transform |
| 7 | Motion Compensation based prediction |
| 8 | Frame Memory |
| 9 | Display |

**Algorithms for Individual Modules:**

**Module 1: Bit-Level Routine**

**Description:** For reading an MPEG-2 video bit stream and extracting each byte this module is implemented.

{**Nomenclature:** test.m2v= tested MPEG-2 file, Bi=input buffer, Bt=temporary buffer, Fb=flush buffer}.

**Begin**

**Step1:** Read MPEG-2 file, test.m2v.

/*This contains the encoded bit stream of MPEG-2 file*/

**Step2:** Initialize input buffer, Bi.

/*Transforming content of MPEG-2 file to this Bi*/

**Step3:** Initialize temp buffer, Bt.

/*Transforming content of Bi to Bt*/

**Step4:** Initialize flush buffer, Fb.

/*Extracting each byte from the Bt, storing them in Fb*/

**Step5:** Stop

**End**

**Module 2: Header Decoding**

**Description:** For extracting the headers of MPEG-2 video file viz., sequence header, group of picture header, picture header and slice header and it's decoding this module is preferred. The extraction of all the parameters of MPEG-2 is done according to the start code values, which is discussed as earlier.

{**Nomenclature:** Bi=input buffer, Sq_L_P=sequence layer parameters, pos=position of buffer level, ld->BntCnt=bit by bit counter, N=total no. of bits in input buffer, n= single bit value in a buffer, G_L_P= group of picture layer parameters, Sl_L_P= slice layer parameters, p_c_t=picture coding type, P=prediction picture, B=bidirectional picture, f_v=forward motion vector, b_v=backward motion vector}

**Begin**

**Step1:** Sequence Header Extraction and its decoding.

/*Important parameters of sequence layer like vertical size, horizontal size, aspect ratio, vbv (variable buffer verifier) delay are extracted from the input buffer, Bi in order to decode the header of sequence layer*/

Sq_L_P=Bi (N); pos=ld->BitCnt;

**Step 1.1:** Perform Zigzag operation on bits

/*By checking the flag of input buffer, the zigzag operation is carried*/

```
if (Bi<0)
  then
    for i=n to N do
  Begin
    • Bit by bit Zig-zag scanning is done until bit 'n' reaches to 'N'
  End
else
   /*No scanning is performed*/
```

**Step2:** Group of Picture Header and Picture Header Extraction and its decoding.

/*Important parameters of group of picture layer like hour, minute, second, DC value are extracted from the input buffer, Bi in order to decode the header of group of picture layer*/

G_L_P=Bi (N); pos=ld->BitCnt;

**Step 2.1:** Assigning the motion vector to pictures.

/*By checking the type of picture and its coding type an appropriate motion vector is assigned*/

```
if ((p_c_t==(P||B)=1))
then
   f_v= Bi(n);
else if (p_c_t==B=1)
```

then

b_v=Bi (n);

else

/*Nothing is performed*/

**Step3:** Slice Header Extraction and its decoding.

/*Important parameters of slice layer like intra frame, slice picture id are extracted from the input buffer, Bi in order to decode the header of slice layer*/

Sl_L_P=Bi (N); pos=ld->BitCnt;

**Step 3.1:** Extraction of Slice Part

/*By checking the flag of input buffer, slice part can be extracted*/

if (Bi(n)<0)

then

- Vertical position of the slice is extracted

else

- Horizontal position of the slice is extracted

**Step4:** Stop

**End**

**Module 3: Macroblock and Motion Vector Decoding**

**Description:** In order to decode the macroblocks and its data this module is used, further motion vector decoding is also discussed.

{**Nomenclature:** Bi=input buffer, Bt_1=temporary buffer1, macb_l_p= macroblock layer parameters, pos=position of buffer level, ld->BntCnt=bit by bit counter, N1=total no. of bits in input buffer, Bt_2 temporary buffer2, fR(x,y)=pixel of reconstructed frame, f1R(x,y)=pixel of estimated image block using an estimated image block, N=size of the image block, V1=variation between current block and that of above block, V2=variation between current block and that of left block, V3=variation between current image block and the one below it}

**Begin**

**Step1:** Extraction of Macroblock header and its data.

/*Macroblock header and its data is extracted bit by bit from the input buffer, Bi and is stored in other buffer, Bt_1*/

**Step2:** Decode Macroblock header.

/*Decoding is carried out with the help of buffer, Bt_1 by considering some of the parameters of Macroblock viz., macroblock type, quantizer code, motion vector forward, motion vector backward according to the predefined bits for these parameters*/

macb_l_p=Bi(N1); pos=ld->BntCnt;

**Step3:** Decode Macroblock data.

/*Decoding of Macroblock data is initialized with help of pointer, the decoded data is transferred to the buffer, Bt_2, now which contains mainly the chrominance and luminance information of decoded blocks*/

**Step4:** Decode of Motion Vectors.

/*Select the motion vector say 'v' among a set of candidate vectors which minimizes the total variation between the boundaries of current image block and those of adjacent ones. The total variation (V [3.4]) is calculated using equations M3.1, M3.2 and M3.3 :*/

$$V1=\sum_{x=x_0}^{N-1+x_0}(f1R(x,y_0)-fR(x,y_0-1))^2 \quad \ldots\ldots(M3.1)$$

$$V2=\sum_{y=y_0}^{N-1+y0}(f1R(x_0,y)-fR(x_0-1,y))^2 \quad \ldots\ldots (M3.2)$$

$$V3=\sum_{x=x_0}^{N-1+x_0}(f1R(x,y_0+N-1)-fR(x,y_0+N))^2 \quad ..(M3.3)$$

V= V1+V2+V3……………………(M3.4)

**Step5:** Store the decoded data.

/*The pixel data and macroblock address are then sent to picture block, which stores the pixel data and corresponding position of the frame*/

**Step6:** Stop

**End**

**Module 4: Variable Length Decoding**

**Description:** Decoding of DC and AC coefficients of the block that are stored in picture block in module3, hence this block is implemented.

{**Nomenclature:** DC=direct current, AC=alternate current, position=current position of the bits, bits_used=no.of bits used, bit_stream=no.of bits read, length=total length of the bitstream, symbol=name assigned to particular bit, I/B/P=Intra/Bidirectional/Predictive pictures}

**Begin**

**Step1:** Maintain the tables to decode intra and non-intra DC coefficients and intra and non-intra AC coefficients.

/*To generate conventional lookup table following procedure is carried out*/

**Step1.1:** Entire Bit stream code is stored in an array.

**Step1.2:** Sorting operation is carried out so that all code words that start with a zero appear first, followed by all code words that starts with a '1'.

**Step1.3:** The source symbols are also stored because the index of the array is no longer is no longer valid after the sort operation.

**Step1.4:** The decoding algorithm follows:

```
position: =0; bits_used: =0;
repeat
    inc (bits_used);
        if bits_stream[bits_used]=1 then
        whilecode[position].bit [bits_used]<>1
  inc (position);
        until bits_used=length[position];
        symbol: =source_symbol [position];
```

**Step2:** Check appropriate picture and motion vector code

/*Use of switch function is made to know the type of picture that is I/B/P picture and type of motion vector code that is forward motion or backward motion in order to decode the corresponding block by assigning the code value from the look up table as discussed in step1*/

**Step3:** Store decoded DC and AC coefficients.

/*The decoded blocks are stored in buffer, B*/

**Step4:** Stop

**End**

**Module 5: Inverse Quantization**

**Description:** In this stage the two dimensional decoded blocks of module 4 are inverse quantized to produce the reconstructed DCT coefficients, hence this block is implemented.

{**Nomenclature:** F=resulting coefficient on the position x and y truncated towards zero, QF=incoming matrix, W=weight matrix or quantization matrix, q_s=quantizer scale defined in picture header}

**Begin**

**Step1:** The DC value is multiplied by a constant, which is specified in the picture header.

**Step1.1** Calculate other coefficients according to the equation M5.1

F [x] [y] = ((2*QF [x][y]+k)*w[x][y]*q_s)/32-----(M5.1)

/*k=0 for intra blocks, k=sign of quantized coefficients for non- intra blocks (1 if quantized (x,y)>0, 0 if quantized (x,y)==0, -1 if quantized (x,y)<0)*/

**Step2:** Store recovered DCT coefficient blocks.

/*The recovered DCT coefficients are stored in the buffer, IQ_B*/

**Step3:** Stop

**End**

**Module 6: Inverse Discrete Cosine Transform**

**Description:** This module performs 8x8 IDCT on decoded DCT coefficient blocks, which are obtained in module 5 in order to get inverse transformed values.

{**Nomenclature:** f (i, j)=IDCT function over two variables i and j (a piece of a block), F (x, y)= function of decoded DCT block over two variable x and y}

**Begin**

**Step1:** Perform row IDCT.

/*Row-wise inverse transformation is carried out on DCT coefficient block*/

for i=0 to N do

Begin

- Performing IDCT on the rows of decoded DCT blocks until the bit 'i' reaches to 'N$^{th}$' block.

End of for 'i'

**Step2:** Perform column IDCT.

/*Column-wise inverse transformation is carried out on DCT coefficient block*/

for j=0 to N do

Begin

- Performing IDCT on the columns of decoded DCT blocks until the bit 'j' reaches to 'N$^{th}$' block.

End of for 'j'

**Step3:** Calculate row and column IDCT according to the equation M6.1

$$f(i,j)=\sum_{x=y=0}^{n}\sum^{n}((c(x)c(y))/4)\ \cos(((2i+1)x\pi)/16)\ \cos(((2j+1)y\pi)/16)\ F(x,y) \ldots\ldots\text{(M6.1)}$$

where: x, y,i, j=0, 1, 2, 3….n

c(x),c(y)=constants= square root of (2)/2 if x,y=0 or c(x),c(y)=1 otherwise.

**Step4:** Stop

**End**

**Module 7: Motion compensation based prediction**

**Description:** Model explains the adding up of predictions from previously decoded pictures with the inverse DCT transformed coefficient data to get the final decoded output.

{**Nomenclature:** p_p=prediction parameters, pos=position of buffer level, ld->BntCnt=bit by bit counter, N=total no. of bits in input buffer, n= single bit value in a buffer, shwtop= show top field, p_c_t=picture coding type, p_t=picture type, predframe= predicted frame, b_ref_frame=backward reference frame, f_ref_frame= forward reference frame, Fi_B= final buffer}

**Begin**

**Step1:** Determination of Predictions

/*Parameters like macroblock type, macroblock motion, picture coding type, frame structure are extracted from buffer, Bt_2 in which the decoded information of macroblocks are stored*/

p_p=Bt_2 (N);      pos=ld->BntCnt;

**Step2:** Determination of frame-based predictions

/*Frame-based predictions are divided into two field's viz., top and bottom fields. Comparing the parameters like forward reference frame (f_r_f), current frame (c_f), and coded picture width (c_p_w) with the parameters listed in step1 the actual prediction is calculated as follows :*/

if (shwtop<n)

then

- Top field is predicted

else

- Bottom field is predicted

**Step3:** Determination of a frame

/*The algorithm chooses some parameters like picture coding type, prediction type, second field, current field which compare these with motion vector forward of a macroblock to calculate correct prediction*/

if ((p_c_t==p_type) && (secondfield)&&(current field)!m_v_f=1)

then

- predframe= b_ref_frame;

else

- predframe= f_ref_frame;

**Step4:** Addition of predictions with IDCT values

/*The predictions which are got in step1 and 2 are added up with IDCT transformed coefficients in order to get final decoded out put, stored in buffer, Fi_B*/

**Step5:** Stop

**End**

**Module 8: Frame Memory**

**Description:** Model is used in order to store the final decoded picture as either one frame or two fields which points the picture in display order.

{**Nomenclature:** P_S= progressive scanning, I_S= Interlaced scanning, Wi_B= Write buffer}

**Begin**

**Step1:** Determination of scanning pattern of a frame.

/*Parameters like progressive sequence (p_s), progressive frame (p_f) and frame store flag (f_s_f) are checked to the know the type of scanning*/

if ((p_s||p_f||f_s_f)==1)

then

- P_S

else

- I_S

**Step2:** Storing of frame in write_buffer, Wi_B

/*By extracting the height and width of the pixel value of the decoded frame, the appropriate storing of the information is done*/

for i=0 to lesser then height do

Begin

- Storing the height wise pixel information in Wi_B until the pointer 'i' reaches the last value

End of for i

for j=0 to lesser then height do

Begin

- Storing the width wise pixel information in Wi_B until the pointer 'j' reaches the last value

End of for j

**Step 3:** Stop

**End**

**Module 9: Display**

**Description:** Model explains the implementation of X11 format in order to display the decoded picture stored in Wi_B. The X11 colors are of 16-bits.

{**Nomenclature:** RGB= red green blue values of the pixel in the blocks}

**Begin**

**Step1:** RGB values of the decoded picture are masked.

**Step2:** Shifting of high bit values of masked picture to get good resolution.

**Step3:** Compare shift values of RGB to know that mask values to be shifted left or right.

**Step4:** Mask values of RGB are assigned to the map entries by taking one variable say cmaplen.

**Step5:** Then this variable is compared with zero in order to fix out the exact values of RGB.

**Step6:** Then at last, the backward and forward values of the pixels are compared with the exact values got in step5 in order to get the entries in direct color map.

**Step7:** Stop

**End**

**Algorithm for Overall MPEG-2 Decoder**

{**Nomenclature:** M_Blocks=Macroblocks, DCT=Discrete Cosine Transform, IDCT=Inverse Discrete Cosine Transform, LUTs=Lookup Tables}

**Begin**

**Step1:** Read MPEG-2 video bit stream by calling **Module1.**

**Step2:** Decode sequence, group of picture, picture and slice headers by calling **Module2.**

**Step3:** Decode M_blocks and motion vectors by calling **Modele3.**

**Step4:** Module 1 and Module 3 are given as inputs to the Variable Length Decoder (**Module 4**) and Inverse Quantization (**Module 5**) successively in order to form DCT coefficients with the help of LUTs.

**Step5:** Perform IDCT on DCT coefficients to get the prediction error blocks by calling **Module 6.**

**Step6:** Perform Motion Compensation (**Module 7**) on prediction blocks to get the final decoded blocks until the last block reaches the final value, other wise repeat the steps 2, 3, 4 and 5 continuously.

**Step7:** Store the decoded blocks in the Frame store by calling **Module 8.**

**Step8:** Display the decoded blocks on to the Displayer by calling **Module 9.**

**Step9:** Stop

End

## IV. RESULTS AND DISCUSSIONS

Figure 4 depicts the Cumulative Bandwidth utilized with respect to (wrt) Time. This graph is drawn by taking seven samples ('x' from x1 to x7) of MPEG-2 video bit streams. Here first 25 frames are taken from each sample with respect to time and corresponding cumulative bandwidth is plotted.

**NOTE**: We observe from figure 4 that in one second first 25 frames are decoded. That is in 40ms first frame is decoded, in 80ms second frame is decoded, in 120ms third frame is decoded, in 160ms fourth frame is decoded and so on. So by this graph we come to know that we can decode 25frames in one second with corresponding bandwidth which is allotted to them.

Figure 5 and figure 6 discuss the Minimum and Maximum Bandwidth Utilized wrt no. of movie songs taken respectively. Here we can observe that the graphs are plotted only by extracting the values of minimum and maximum frame bandwidth of each sample which are taken in to account in order to draw figure 4 respectively. The curves in the figure 5 and figure 6 show the variation for number of movie samples with fixed requirement of bandwidth. This performance parameter curve will be very useful in order to get the minimum and maximum bandwidth utilization from available bandwidth respectively. So from the figure 5 and figure 6 we notice that a minimum of 12200 bits and maximum of 17907 bits of bandwidth is allotted in order to reconstruct the original videos which are taken as an example.

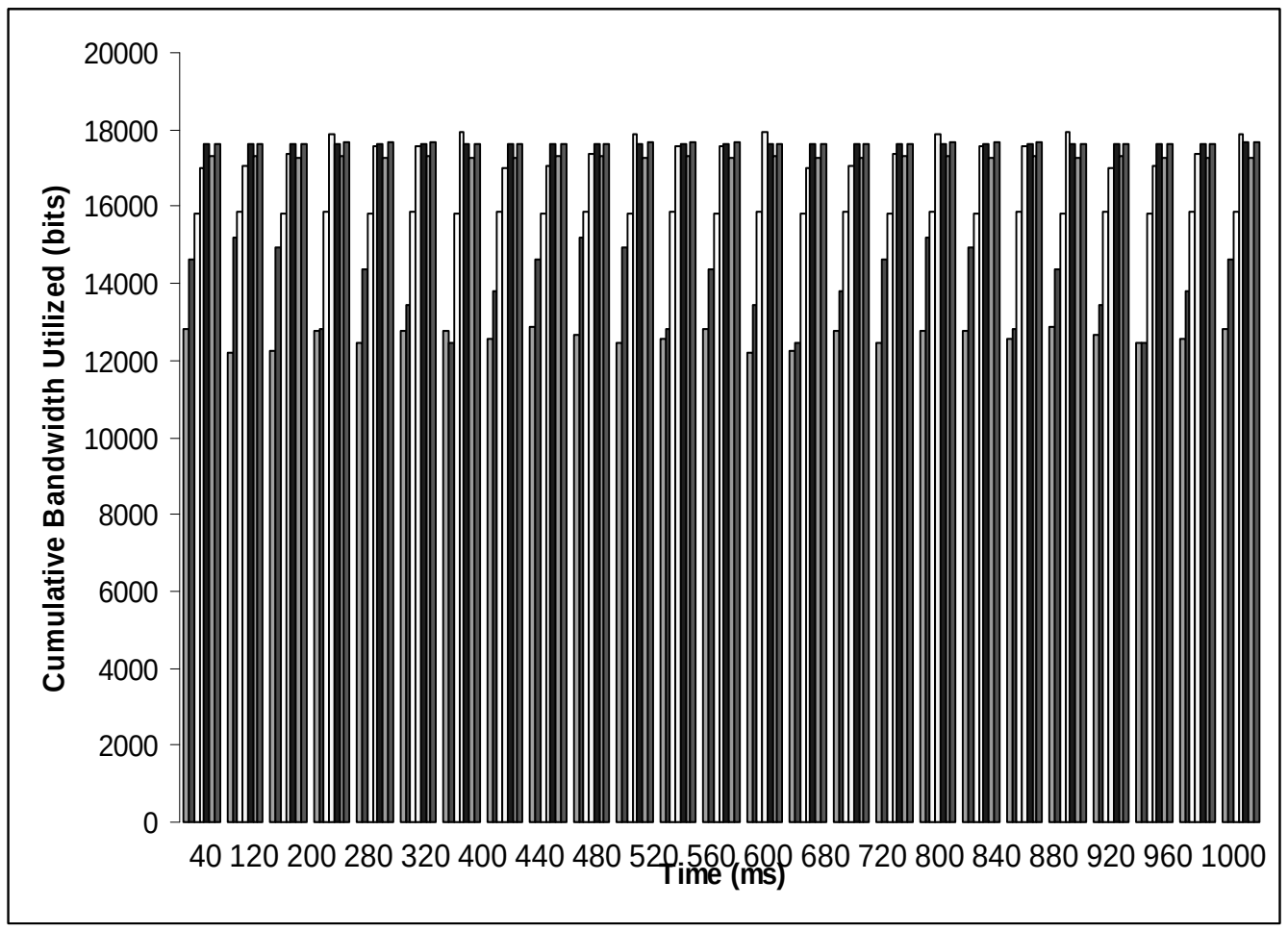

Figure 4: Cumulative Bandwidth utilized wrt to time

Figure 7 depicts the Average Bandwidth Utilized wrt no. of movie songs taken. This graph is drawn only by extracting the values of average frame bandwidth of each sample which are taken in figure 4.

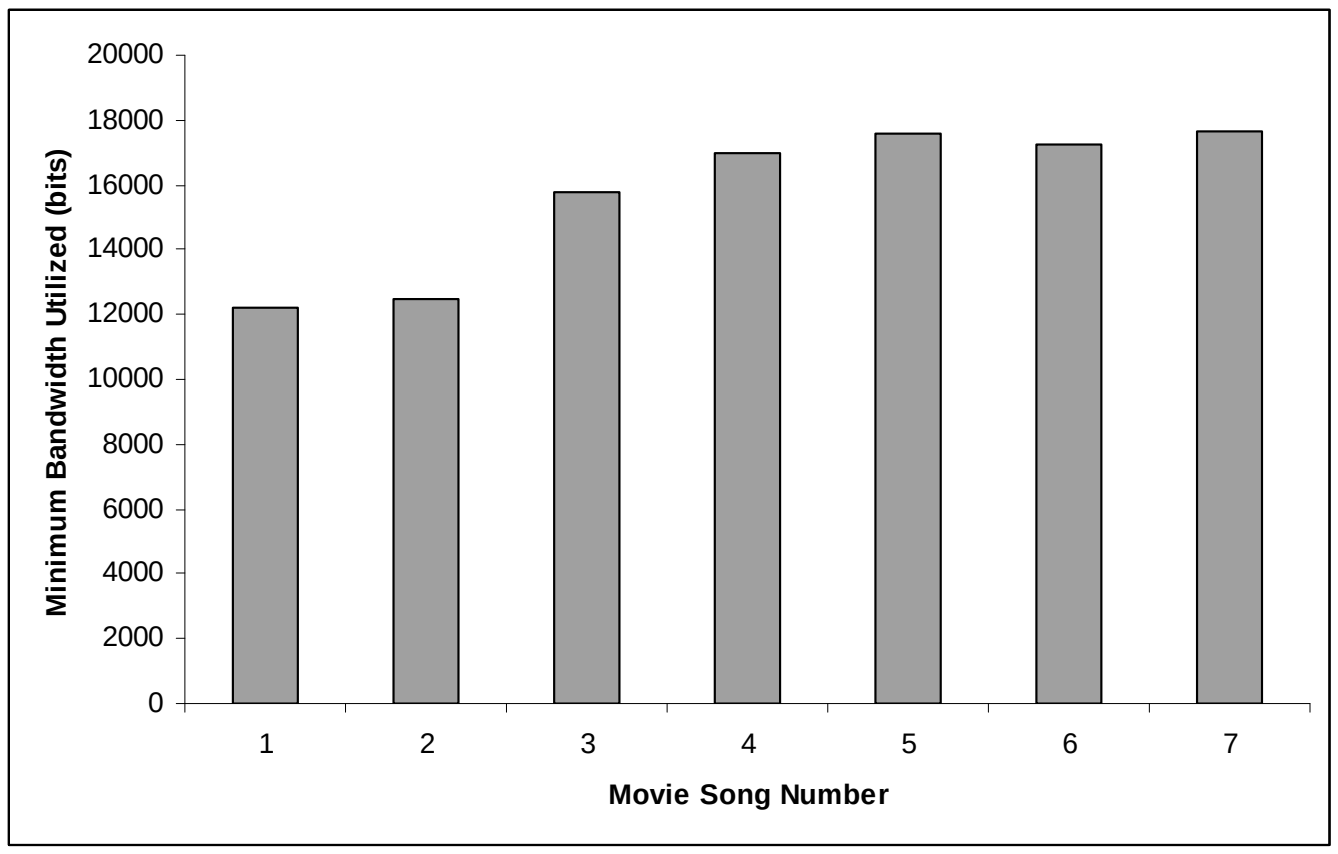

Figure 5: Minimum Bandwidth Utilized wrt movie song number

Here the curve in the graph shows the variations for number of movie clips with fixed requirement of bandwidth. This performance parameter curve will be very useful in order to get the average bandwidth utilization from available bandwidth. So from the figure 7 we can deduce that a average of 16063 bits of bandwidth is allotted in order to reconstruct the original videos which are taken as an example.

## CONCLUSION

In this work some of the performance parameters analyzed for MPEG 2 decoder such as: Cumulative Bandwidth, minimum bandwidth, maximum bandwidth and average bandwidth. The scheme has been developed on UNIX platform with one concern that since of all MPEG standards deal with compression only MPEG-2 addresses the transmission, or movement, of compressed digital content across a network, and UNIX is preferred for networking applications. Finally we can conclude that the MPEG-2 decoder which is designed is flexible, robust, maintainable and effective bandwidth utilization.

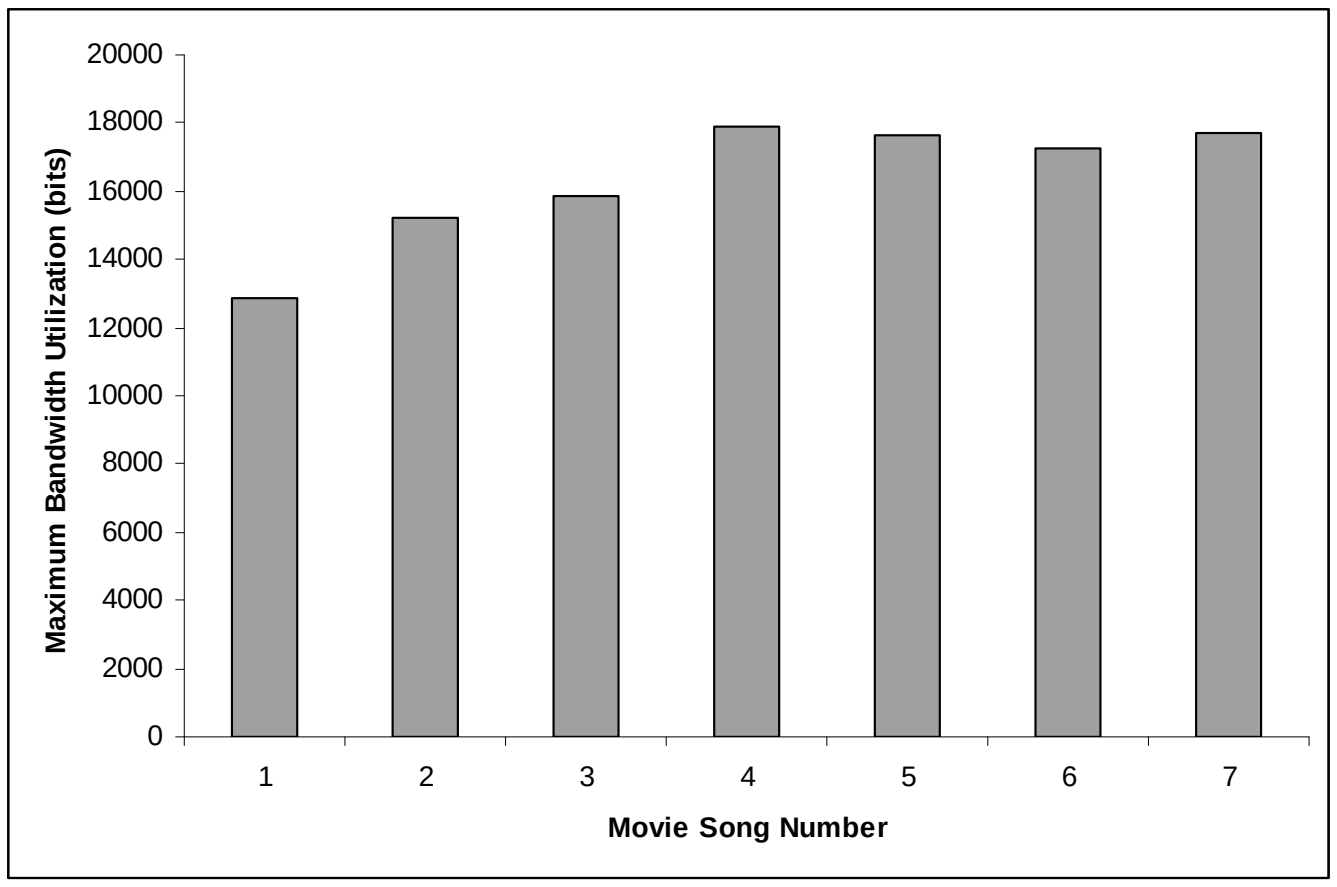

Figure 6: Maximum Bandwidth Utilized wrt movie song number

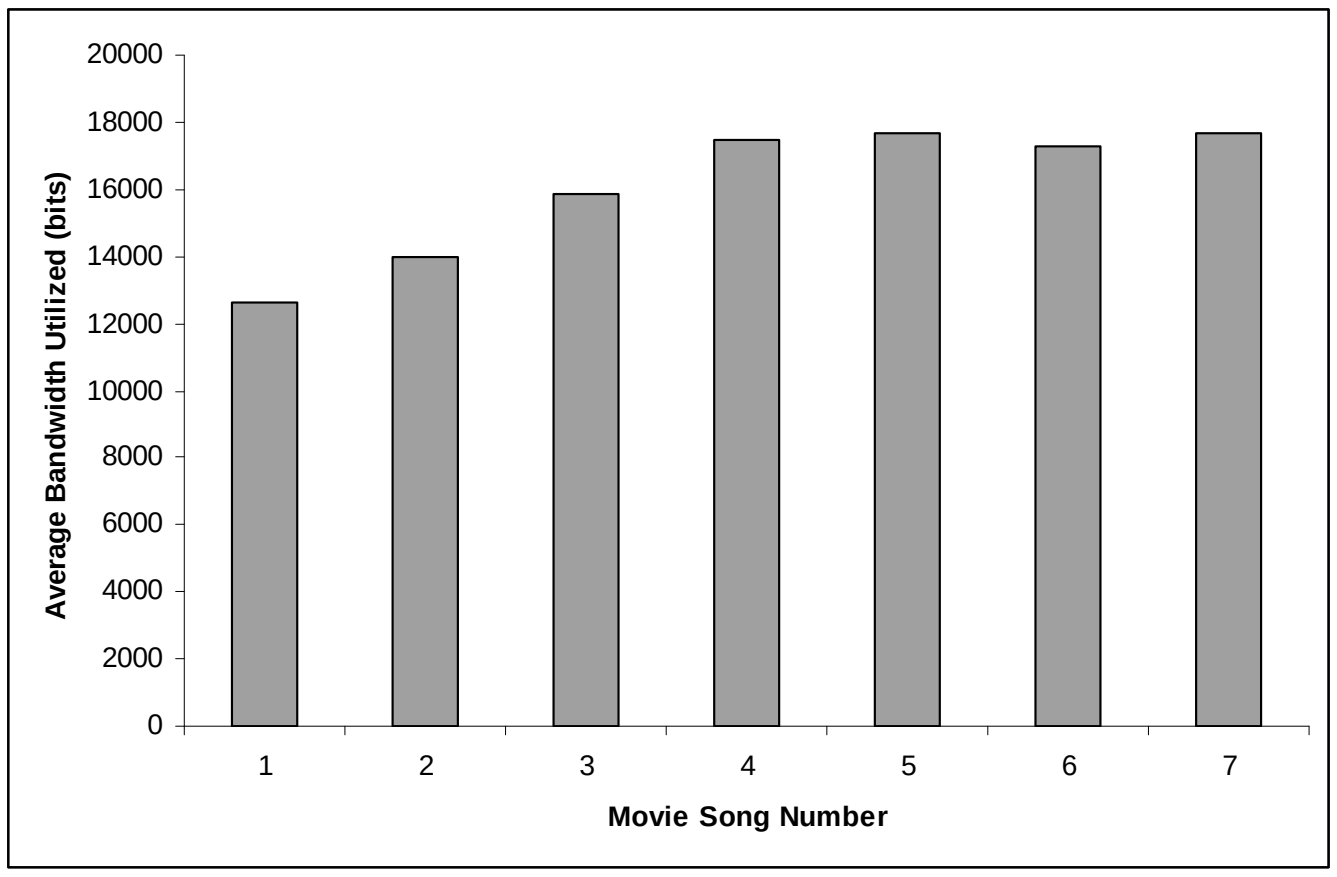

Figure 7: Average Bandwidth Utilized wrt movie song number

AUTHORS PROFILE

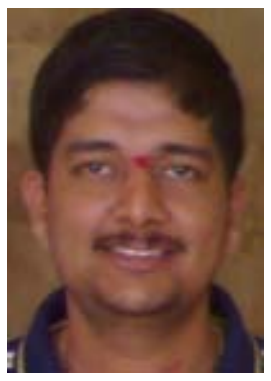

**SANDEEP. KUGALI** completed his M.Tech from Visvesvaraya Technological University Belgaum, Karnataka. Presently he is serving as a Lecturer of Department of Information Science and Engineering Bagalkot, Karnataka. His areas of interest include C- programming, Computer Networks, Unix Shell Programming and Multimedia Communications. He has published 01 paper in referred National Conference.

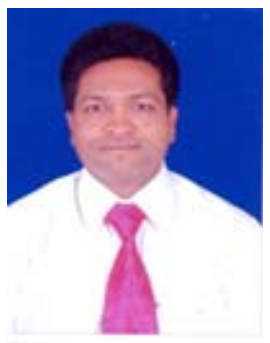

**S. S. Manvi** completed his PhD from Indian Institute of Science, Bangalore. Presently he is serving as a Professor of Electronics and Communication Engineering, REVA Institute of Technology and Management, Bangalore. His areas of interest include wireless/wired network, AI applications in network management, Grid Computing and Multimedia Communications. He has published over 25 papers in referred National/International Journals and 60 papers in referred National/ International Conferences.

He has coauthored books "Communication Protocol Engineering" and "Computer Concepts and C Programming" published by PHI. He is a reviewer of several national/international journals. He is a member of IEEE USA, Fellow of IETE, India Fellow of IE, India and member of ISTE, India. He has been included in Marqui's Who's who in world and International Biographies of Cambridge, London in the year 2006.

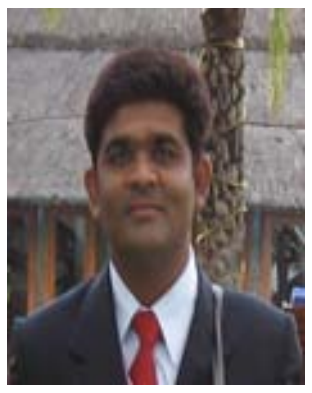

**A. V. SUTAGUNDAR** completed his M. Tech from Visvesvaraya Technological University Belgaum, Karnataka. He is pursing his PhD in the area of Cognitive Agent based Information Management in wireless Networks. Presently he is serving as an Assistant Professor of Department of Electronics and Communication Engineering Bagalkot, Karnataka. His areas of interest include Signal and system, Digital Signal Processing, Digital Image Processing, Multimedia Networks, Computer communication networks, Wireless networks, Mobile ad-hoc networks, Agent technology. He has published 15 papers in referred National/International Conferences.